\documentclass[fl eqn,twoside]{article}
\usepackage{amsmath}
\usepackage{amsfonts}
\usepackage{amssymb}
\usepackage{graphicx}
      \newcommand{\cl}{\mathcal{L}}

\def\wrt{with respect to \ }

\def\be{\begin{equation}}
\def\ee{\end{equation}}
\def\bi{\bibitem}

\begin{document}
\title{Canonical formulation of scalar curvature squared action in higher dimensions}
\author{Subhra Debnath $^{\dag}$,  Soumendranath Ruz $^{\ddagger}$ and Abhik Kumar Sanyal $^{*}$}
\maketitle

\noindent

\begin{center}
$^{\dag}$, $^{*}$ Dept. of Physics, Jangipur College, Murshidabad, India - 742213\\
$^{\ddagger}$ Dept. of Physics, University of Kalyani, Nadia, India - 741235\\
\end{center}

\footnotetext[1]{\noindent
Electronic address:\\
\noindent $^{\dag}$subhra\_ dbnth@yahoo.com\\
\noindent $^{\ddagger}$ruzfromju@gmail.com\\
\noindent $^{*}$sanyal\_ ak@yahoo.com\\}
\begin{abstract}

\noindent Canonical formulation for an action containing scalar curvature squared term $(R^2)$ in arbitrary dimension has been performed in maximally symmetric space-time. The quantum dynamics does not alter significantly from the same in $4$-dimension. Classical solution is also at par with the one presented by Starobinsky. WKB approximation peaks around the classical solution.
\end{abstract}

\section{Introduction}
General theory of Relativity is plagued with an ultraviolet catastrophe being typically manifested in cosmological or black-hole type singularities. Any resolution to this problem requires a theory which is well behaved in the UV region and reduces suitably to Einstein's gravity in the infrared region. It was Stelle \cite{Stelle} who first proved that an action in the form
$A = \int d^4x\sqrt{-g}\big[{R\over 16\pi G} + \beta R^2 + \gamma R_{\mu\nu}R^{\mu\nu}\big]$
is renormalizable, since it leads to a graviton propagator which behaves like \textrm{k}$^{-4}$ for large momenta. Further Newtonian limit of the static field for the above action is
$h_{00} = {1\over r} + {1\over 3}{e^{-m_0 r}\over r} - {4\over 3}{e^{-m_2 r}\over r}$
where, $m_0 = [32\pi G(3\beta - \gamma)]^{-{1\over 2}}$ and $m_2 = (16\pi G \gamma)^{-{1\over 2}}$. In the weak field limit, $h_{00}$ approaches Newtonian limit $r^{-1}$ by ensuring $m_0$ and $m_2$ large enough. The action is asymptotically free \cite{Tomb, Frad}, ie., interaction becomes very weak at arbitrarily large energy scale. Despite such wonderful features, the theory falls short of the fact that it contains ghost degrees of freedom when expanded in the perturbative series about the linearized theory. Apart from the two familiar massless spin-2 gravitons arising out of the linearized field energies of these particle excitations, there exists one massive scalar particle, and five massive spin-2 particles. The linearized energies of the massive spin-2 excitations is negative definite. It is possible to cast the theory so that the massive spin-2 eigenstates of the free field Hamiltonian has positive definite energy, but in the process, negative norm in the state vector space emerges, which destroys the unitarity of the S-matrix \cite{Pais}. With $\gamma = 0$, the ghost degree of freedoms disappear, but the ultraviolet catastrophe reappears. So, it is clear that although the scalar curvature squared term ($R^2$) does not improve the ultraviolet behaviour, it is in no way responsible for the violation of unitarity of the theory.\\

No attempt to formulate renormalized quantum theory of gravity could avoid the presence of $R^2$ term from the action. For example, string generated gravity models \cite{Boul1}, string inspired theory of gravity \cite{String}, Lovelock gravity \cite{Love}, four-dimensional brane world effective action \cite{Brane} along with the Ho\v{r}ava-Lifshitz gravity \cite{Horava} also contain this term. Now, in the absence of a complete theory of gravity, quantum cosmology was initiated in an anticipation that it might possibly extract new physics leading to a path towards quantum gravity. Therefore, to study the issue of quantum cosmology, it is required to cast the action containing $R^2$ term in canonical form. A host of canonical formalisms of higher order theory of gravity appears in the literature \cite{Ostro} - \cite{HL}. However, even for an action containing $(\alpha R +\beta R^2)$ term, all the formalisms either produce modified Wheeler-DeWitt equation which doesn't provide standard quantum mechanical probabilistic interpretation or produce Schr\"odinger-like equation but suffer from the disease of the loss of unitarity. This problem was circumvented by Sanyal and Modak \cite{Sanyal1} and Sanyal \cite{Sanyal2} by choosing auxiliary variable, required for canonical formulation, judiciously. In the Robertson-Walker(RW) minisuperspace model containing lapse function $N(t)$, it was possible to cast the above action in the canonical form \cite{Sanyal3} $A = \int[\dot h_{ij}p^{ij} + \dot K_{ij}\Pi^{ij} - N\mathcal{H}] dt d^3 x,$
where, $h_{ij}$ and $K_{ij}$ are the metric on three space and the extrinsic curvature tensor, $p^{ij}$ and $\Pi^{ij}$ are their canonically conjugate momenta respectively. $\mathcal H$ is the constrained Hamiltonian. It was noticed that only a particular choice of auxiliary variable leads to a viable quantum cosmological model, since it yields Schr\"odinger-like equation for which the effective Hamiltonian turned out to be hermitian. Thus the standard form of continuity equation and the quantum mechanical probabilistic interpretation are admissible. Classical field equations were found to admit solutions obtained by Starobinsky \cite{Staro}. Under WKB approximation the wavefunction turned out to be oscillatory being peaked around a classical inflationary solution. However, from the standpoint of dimensional regularization, it is important to quantize $R^2$ gravity in spaces of higher dimension. In this regard, it is required to formulate the theory of $R^2$ gravity for $D > 4$, and a constituent part of this formulation is canonical quantization. A model of multidimensional $R^2$ gravity finds natural expression in Kaluza-Klein approach, in which a central part is played by the geometry of space with $D > 4$, both at the classical and the quantum levels \cite{Kk}. In addition, multidimensional $R^2$ gravity follows from superstring theory in $D = 10$ \cite{St} as an effective theory corresponding to the Planck's energy. It is therefore of interest to construct quantum cosmology in the framework of multidimensional $R^2$ gravity. In view of the nice features already realized in $D = 4$, we proceed here to perform canonical formulation of $R^2$ theory of gravity in arbitrary higher dimensions. In this context, we would like to mention that $R_{\mu\nu}^2$ term has not been taken into consideration here, because an unique boundary term required to supplement the action has not been found yet. Further, in RW minisuperspace, $R_{\mu\nu}R^{\mu\nu} - {D\over 4(D-1)}R^2$ is topologically invariant, and so $R_{\mu\nu}^2$ term can always be replaced by $R^2$ term.\\

In the following section, we take up the action and briefly review the issue of boundary terms required to supplement the action under consideration. In section 3, we follow our earlier technique \cite{Sanyal3} to formulate canonical action and proceed to make WKB approximation. We conclude in section 4.
\section{The issue of boundary term}

In case of gravity, no nontrivial lagrangian $\mathcal{L}_g$ can be constructed from the metric $g_{\mu\nu}$ and its first derivatives alone. Rather, even for the Einstein-Hilbert action, the Lagrangian depends on second derivatives of the metric, although only linearly, in the form, $\mathcal{L}_g[g, \partial g, \partial^2 g]$. To construct the canonical action, it is required to integrate the Lagrangian over a spacetime volume $\mathcal{V}$, and the second derivative term gives a contribution only on the boundary $\partial\mathcal{V}$. Under metric variation, setting $\delta g_{\mu\nu}\Big{|}_{\partial\mathcal{V}} = 0$ as usual, the Einstein-Hilbert action in $D$ dimension yields
\be \label{26} \delta A = \frac{1}{16\pi G}\int_{\mathcal{V}} d^D x\sqrt {-g}\left(R_{\mu\nu} - \frac{1}{2}R g_{\mu\nu} \right)\delta g^{\mu\nu} + \frac{1}{16\pi G}\oint_{\partial \mathcal{V}} d^{(D-1)}x \sqrt h h^{\mu\nu}\partial_\sigma(\delta g_{\mu\nu})n^\sigma .
\ee
Since the resulting boundary term contains derivative of the metric, it can not be set equal to zero. Rather, one can add a term to the Einstein-Hilbert action, such that its variation cancels out this term. The variation of
\be\label{s1} \Sigma_{R_D} = \frac{1}{8\pi G}\oint_{\partial \mathcal{V}} {d^{(D-1)} x} \sqrt{h}~K \ee
gives the second term of equation (1) \cite{York}. Here, $K$ is the trace of $K_{ij}$. Thus, Einstein-Hilbert action must be supplemented by the so called Gibbons-Hawking-York boundary term (2), and the complete gravitational action for general relativity should be expressed as,
\be\label{EH2} A_E = \frac{1}{16\pi G}\int_{\mathcal{V}}\sqrt{-g}d^D x R + \frac{1}{8\pi G}\oint_{\partial \mathcal{V}} {d^{(D-1)} x} \sqrt{h}~K. \ee
Canonical formulation of Einstein-Hilbert action was presented by Arnwitt, Deser and Misner (ADM) \cite{Arnwitt}, highlighting the importance of the constraints and showing that the Hamiltonian is precisely the spatial integral of the constraints. However, inadvertently they discarded the total derivative term inherently present in the gravitational action. On the contrary, De-Witt kept the total derivative term, which becomes a surface term at spatial infinity. In the process, he recovered the fact that the Hamiltonian coincides with the ADM energy \cite{DeWitt}. This fact unveils the importance of boundary term. He also formulated quantum version of Einstein gravity which is known as Wheeler-DeWitt Equation.\\

\noindent
Likewise, higher order theory of gravity $f(R) \propto R^n$ should also be supplemented by appropriate boundary term. Under metric variation, the action,
\be\label{f(R)} S = \int d^D  x\sqrt{-g}f(R), \ee
can be expressed in the form \cite{Madsen}
\be \delta S = \int_{\mathcal{V}} {d^D x}\sqrt{-g}\left[ \Big(R_{\mu\nu} + g_{\mu\nu}\Box  - \nabla_\mu\nabla_\nu\Big) f'(R) - \frac{1}{2}g_{\mu\nu}f(R) \right]{\delta g^{\mu\nu}} + \oint_{\partial\mathcal{V}} d^{(D-1)}x\sqrt{h}f'(R) h^{\mu\nu}\partial_\sigma(\delta g_{\mu\nu})n^\sigma. \ee
For $f(R) \propto R$, the surface term appearing in (5) coincides with the one appearing in (1). However, for $f(R)$ other than $R$, second term of the above equation does no cancel with Gibbons-Hawking-York like boundary term under the only assumption that $\delta g_{\mu\nu}\Big{|}_{\partial\mathcal{V}} = 0$. Rather, some extra condition is required. Barth \cite{Barth} and Horowitz \cite{Horowitz} had chosen the first normal derivative of $g$ (second fundamental form) to be fixed on the boundary. So there is no need to add any boundary term. But it restricts the wide generality of the solutions \cite{Dyer}. On the contrary, the variation of $2\oint_{\partial\mathcal{V}} \sqrt h f'(R)Kd^{(D-1)} x$
produces the surface term appearing in equation (5) under the additional condition $ \delta R\Big{|}_{\partial\mathcal{V}} = 0 $ at the surface. Therefore the action under consideration here,
\be\label{AD} A = \int \left[\frac{R - 2\Lambda}{16\pi G} + \beta R^2\right]\sqrt {-g} d^D x\ee
should be supplemented by the following boundary term
\be\label{Boun}\Sigma = \Sigma_{R_D} + \Sigma_{R^2_D} = \frac{1}{8\pi G}\int \sqrt h Kd^{(D-1)}x + 4\beta \int \sqrt h K Rd^{(D-1)}x.\ee
Our present aim is to perform canonical formulation of action (\ref{AD}) being supplemented by the boundary term (\ref{Boun}) in the $D$ dimensional RW minisuperspace model
\be\label{RW} ds^2 = -N^2 dt^2 + a(t)^2\left[{dr^2\over 1-kr^2} + r^2 d\theta^2 + r^2 sin^2\theta d\phi^2 +d X_{\delta}^2\right],\ee
where, $N = N(t)$ is the lapse function and the extra dimensions $\delta = D - 4$. The expression of Ricci scalar is,
\be\label{R}
R = {(D-1)\over N^2}\left[2{\ddot a\over a} + (D-2){\dot a^2\over a^2}- 2{\dot a\dot N \over a N}\right] + 6\frac{k}{a^2}.
\ee
As already mentioned in the introduction, a host of canonical formalisms corresponding to higher order theory of gravity appears in the literature. Ostrogradski's technique comes to the first place \cite{Ostro}. The underlying idea of this method and of its subsequent generalizations (constrained system was included by Gitman et. al. \cite{Gitman}, being modified by Buchbinder and his co-others \cite{BL} while the occurrence of constraints at the Lagrangian level were precluded by a modified technique developed by Schmidt \cite{Schmidt}) consist in introducing, besides the original configuration space variables, a new set of coordinates that encompasses each of the successive time derivatives of the original coordinates so that initial higher-order regular system be reduced to a first-order system. However, when applied in higher order theory of gravity (\ref{AD}) in RW minisuperspace (\ref{RW}), Schr\"odinger-like equation is produced, but the effective Hamiltonian operator is not hermitian. Even serious problem is that, these techniques do not take boundary term (\ref{Boun}) into account and therefore treat a Lagrangian in the form $\mathcal{L} = \mathcal{L}(x,\dot x,\ddot x)$ and hence Einstein-Hilbert action as higher order theory. In the process, it yields trivial solution for scalar-tensor theory of gravity. Therefore Ostrogradski's technique along with its generalizations are not suitable for the theory of gravity.\\

Boulware \cite{Boul2} on the contrary, presented an elegant technique of canonical formulation for the whole superspace, identical to the ADM formulation of General Relativity \cite{Arnwitt}, taking the most general form of the quadratic gravitational action and the corresponding  boundary term into account. Apart from the basic phase space variables $(h_{ij}, p^{ij})$, Boulware took another set $(Q_{ij}, P^{ij})$, where the auxiliary variable $(Q_{ij})$ is obtained by varying the action \wrt the highest Lie derivative of $K_{ij}$ as $Q^{ij} = -{\partial\cl\over\partial(\cl_n K_{ij})}$, present in the action and taking only that part from it which vanishes in the flat space. Thus in the process of choosing auxiliary variable one does not pick out a part from the Linear term. The momentum, canonically conjugate to $Q_{ij}$ is simply the extrinsic curvature tensor $P^{ij} = K^{ij} = -{1\over 2}\cl_n g_{ij}$. However, when applied to the action (\ref{AD}) in the RW minisuperspace model (\ref{RW}), one obtains a modified version of Wheeler-DeWitt equation. Thus the technique does not admit standard quantum mechanical probabilistic interpretation. Further, the quantum dynamics presented in terms of the auxiliary variable ($Q = a^3 R$) is of no practical use. To get around the difficulty, Horowitz \cite{Horowitz} in contrast, chosen an auxiliary variable by varying negative of the action with respect to the highest derivative of the field variable present in the action. The Hamiltonian was then expressed in terms of basic variables ($h_{ij}, p^{ij}, K_{ij}, P^{ij}$). In the process a Schr\"odinger-like equation is arrived at. Nevertheless the effective Hamiltonian is non-hermitian. Further, the auxiliary variable picks up a part also from the linear term unlike Boulware's method. Therefore such technique again considers a Lagrangian in the form $\mathcal{L} = \mathcal{L}(x,\dot x,\ddot x)$ to be higher order theory. Hence the method may be applied for scalar-tensor theory of gravity as well, restricting the solution by and large, since the Ricci scalar vanishes $R =0$ in view of the field equation \cite{Sanyal3, Pollock}. \\

Vilenkin \cite{Vilenkin} on the other hand proposed that for a particular minisuperspace, treating $R$ as a constraint of the theory one can introduce it in the action through a Lagrange multiplier $(\lambda)$ and consequently cast the action corresponding to an arbitrary $f(R) \propto R^n$ theory of gravity, in canonical form. Expressing the action (\ref{f(R)}) as,
\be A = \int \left[f(R) + \lambda \left\{R -6\left({\ddot a\over a}+{\dot a^2\over a^2}+{k\over a^2}\right)\right\}\right]\sqrt {-g} d^4x\ee
and varying \wrt $R$ $\lambda$ is found. When substituted in the action, it turns out to be canonical with the variables $a$ and $R$. The corresponding canonical point Lagrangian reads,
\be \cl = a^3(f-f_{,R}R) + 6a^2f_{,RR}\dot R\dot a + 6f_{,R}a\dot a^2 - 6kaf_{,R}. \ee
Canonical quantization leads to
\be \frac{\hbar^2f_{,R}}{6a^3f^2_{,RR}}\frac{\partial^2\psi}{\partial R^2} -\frac{\hbar^2}{6a^2f_{,RR}}\frac{\partial^2\psi}{\partial a\partial R} -[a^3(f-f_{,R}R) - 6kaf_{,R}]\psi = 0. \ee
Identical form was also obtained by Kasper \cite{Kasper} through scalar-tensor equivalence without invoking conformal transformation. In this technique appropriate boundary term (\ref{Boun}) had been accounted for. The underlying reason for the simplicity of the technique lies in the fact that the scalar curvature $R$ has been associated as a variable of the theory. This is forbidden, since as already mentioned, for higher order theory of gravity $\delta R$ must vanish at the boundary. This means $R$ has been treated as a real variable instead of an auxiliary one. While, the variation of the action \wrt $R$ yields nought. Finally, one does not obtain a Schr\"odinger-like equation and so the probabilistic interpretation remains obscure as well. Note that Hawking and Luttrell \cite{HL} did the job by choosing an auxiliary variable in the form $Q = a(1+2\beta R)$ treating $a$ and $R$ to be independent variables as well \footnote{Note that the surface term contains $2\int d^3 x \sqrt{h}[K(1+2\beta R)]$}. The idea behind such a choice is that, under conformal transformation $\tilde{g}_{\mu\nu} = (1+2\beta R) g_{\mu\nu}$, scalar-tensor equivalence is established. In the process, Wheeler-DeWitt equation resembles with scalar-tensor theory of gravity. This technique is restricted to curvature squared gravity only and suffers from the disease already mentioned.\\

It was Sanyal and Modak \cite{Sanyal1}, who first circumvented the problem following a unique scheme, stating that the apart from the boundary term corresponding to linear sector (Einstein-Hilbert action) a part of the boundary term $\Sigma_{R^2_{D1}} = 4\beta\int ({^3R})K\sqrt{h} d^{(D-1)} x$, corresponding to $R^2$ term of the action should be taken care of, prior to the inclusion of auxiliary variable. The technique was followed up by Sanyal \cite{Sanyal2} and his collaborators \cite{Sanyal3}, who finally presented the following list of steps required for the purpose.

\begin{itemize}
  \item Express the action in terms of $h_{ij}$ and remove possible total derivative terms which cancel $\Sigma_{R_{D}}$ and $\Sigma_{R^2_{D1}}$.
  \item Introduce auxiliary variable in the action following Horowitz's proposal \cite{Horowitz}. Integration by parts then takes care of rest of the boundary terms ($\Sigma_{R^2_{D2}} = 4\beta\int({^4R-^3R})K\sqrt{h} d^{(D-1)} x$). The action is then automatically expressed in canonical form.
  \item Cast Hamiltonian constraint equation in terms of basic variables and quantize.
\end{itemize}
In the process, one obtains Schr\"odinger-like equation, where an internal variable acts as time parameter and the effective Hamiltonian becomes hermitian. The fact that unitarity of a renormalized theory of gravity by no means is hampered by the presence of $R^2$ term, signals our proposition is correct. In the following section we extend our work in $D$-dimension.

\section{Canonical formulation of $R^2$ in higher dimension}
In view of the glimpse of earlier works in the context of canonical formulation of scalar curvature squared action, here we proceed, following Sanyal \cite{Sanyal3} to do the same for the action underneath in $D$ dimension.
\be\label{A4} A = \int \left[{R - 2\Lambda \over 16\pi G} + \beta R^2\right]\sqrt{-g}\;d^Dx + \Sigma_{{R}_{D}} + \Sigma_{{R^2}_{D1}} + \Sigma_{{R^2}_{D2}},\ee
where $\Sigma_{{R}_{D}}$, $\Sigma_{{R^2}_{D1}} = 4\beta\int {^3R} K {\sqrt h}~d^{(D-1)}x$, $\Sigma_{{R^2}_{D2}} = 4\beta\int({^4R}-{^3R})K{\sqrt h}~d^{(D-1)}x$ are the corresponding boundary terms. As already mentioned, to account for the boundary terms appearing in (\ref{A4}) we choose a variable $z=a^{\frac{D}{2}} = {h_{ij}}^{\frac{D}{4}}$ and express the action (\ref{A4}) as,
\be\begin{split}\label{37} A & = \int \Big[{1 \over 8\pi G}\Big\{{2(D-1) \over DN} z^{D-2 \over D}\Big(\ddot z - {\dot N \dot z \over N}\Big)  + 3kN z^{2(D-3) \over D} - \Lambda N z^{2(D-1) \over D}\Big\} + 4\beta\Big\{4\frac {(D-1)^2}{D^2 N^3} ~z^{-\frac{2}{D}} \Big({\ddot z}^2 - 2 \frac{\dot N}{N} \dot z \ddot z\\
& + \frac{{\dot N}^2}{N^2}{\dot z}^2\Big) + 12k \frac{(D-1)}{DN} ~z^{\frac{(D-6)}{D}} \Big(\ddot z - \frac{\dot N}{N} \dot z\Big) + 9 k^2 N z^{\frac{2(D-5)}{D}}\Big\}\Big]dt + \Sigma_{{R}_{D}} + \Sigma_{{R^2}_{D1}} + \Sigma_{{R^2}_{D2}}, \end{split}\ee
\noindent
apart from a constant term arising out of the integration over the space part. Now  under integration by parts, some of the total derivative terms viz. ($\Sigma_{{R}_{D}}$ and $\Sigma_{{R^2}_{D1}}$) get cancelled with the boundary terms and in the process the action (\ref{37}) reduces to
\be\label{38}\begin{split} A & = \int \Big[{1 \over 8\pi G}\left\{-2{(D-1)(D-2) \over D^2N} z^{-{2 \over D}}\dot z^2 + 3kN z^{2(D-3) \over D} - \Lambda N z^{2(D-1) \over D}\right\} + 4\beta \Big\{4\frac {(D-1)^2}{D^2 N^3} ~z^{-\frac{2}{D}} \Big({\ddot z}^2 - 2 \frac{\dot N}{N} \dot z \ddot z\\
& + \frac{{\dot N}^2}{N^2}{\dot z}^2\Big) - 12k \frac{(D-1)(D-6)}{D^2 N} ~z^{-\frac{6}{D}} ~\dot z^2 + 9 k^2 N z^{\frac{2(D-5)}{D}}\Big\}\Big]dt + \Sigma_{{R^2}_{D2}}. \end{split}\ee
Now introducing the auxiliary variable following Horowitz \cite{Horowitz} at this stage
\be\label{Q4} Q = \frac{\partial A}{\partial \ddot z} = 32\beta\frac{(D-1)^2}{D^2N^3 } ~z^{-\frac{2}{D}} \Big(\ddot z  - \frac{\dot N }{N} \dot z\Big),\ee
the action (\ref{38}) may be expressed judiciously as,
\be\label{A4b}\begin{split} A &= \int\Big[{1 \over 8\pi G}\Big\{-2{(D-1)(D-2) \over D^2N} z^{-{2 \over D}}\dot z^2 + 3kN z^{2(D-3) \over D} - \Lambda N z^{2(D-1) \over D}\Big\} + Q \ddot z - \frac{\dot N}{N}\dot z Q\\
& - \frac{D^2 N^3 }{64 \beta (D-1)^2} ~z^{\frac{2}{D}} Q^2 - 48\beta k \frac{(D-1)(D-6)}{D^2 N} ~z^{-\frac{6}{D}} ~\dot z^2 + 36\beta k^2 N z^{\frac{2(D-5)}{D}}\Big]dt + \Sigma_{{R^2}_{D2}},   \end{split}\ee
\noindent
so that after removing rest of the total derivative terms, the action in its final canonical form is expressed as,
\be\begin{split} A &= \int\Big[{1 \over 8\pi G}\left\{-2{(D-1)(D-2) \over D^2N} z^{-{2 \over D}}\dot z^2 + 3kN z^{2(D-3) \over D} - \Lambda N z^{2(D-1) \over D}\right\} - \dot Q \dot z - \frac{\dot N}{N}\dot z Q\\
& - \frac{D^2 N^3 }{64 \beta (D-1)^2} ~z^{\frac{2}{D}} ~Q^2 - 48\beta k \frac{(D-1)(D-6)}{D^2 N} ~z^{-\frac{6}{D}} ~\dot z^2 + 36\beta k^2 N z^{\frac{2(D-5)}{D}}\Big]dt. \end{split}\ee
The canonical momenta are,
\be\begin{split}\label{pz5} p_Q &= - \dot z, \;\;\;\;p_N = -\frac{Q \dot z}{N}\\
 p_z & = -{(D-1)(D-2) \over 2\pi G D^2N} z^{-{2 \over D}}\dot z - \dot Q - \frac{Q \dot N}{N} - 96\beta k \frac{(D-1)(D-6)}{D^2 N} ~z^{-\frac{6}{D}} ~\dot z.\end{split}\ee
The $N$ variation equation is
\be\label{45}\begin{split} & - {1 \over 8\pi G}\left\{2{(D-1)(D-2) \over D^2N^2} z^{-{2 \over D}}\dot z^2 + 3k z^{2(D-3) \over D} - \Lambda z^{2(D-1) \over D}\right\} - \frac{Q \ddot z}{N}- \frac{\dot Q \dot z}{N} + \frac{3D^2 N^2 }{64 \beta (D-1)^2} ~z^{\frac{2}{D}} ~Q^2\\
& - 48\beta k \frac{(D-1)(D-6)}{D^2 N^2} ~z^{-\frac{6}{D}} ~\dot z^2 - 36\beta k^2 z^{\frac{2(D-5)}{D}} = 0.\end{split}\ee
Now removing $\ddot z$ term in view of the definition of $Q$ given in (\ref{Q4}), equation (\ref{45}) is expressed as,
\be\begin{split} &- {1 \over 8\pi G}\left\{2{(D-1)(D-2) \over D^2N^2} z^{-{2 \over D}}\dot z^2 + 3k z^{2(D-3) \over D} - \Lambda z^{2(D-1) \over D}\right\} - \frac{\dot Q \dot z}{N} - \frac{\dot N}{N^2} Q \dot z + \frac{D^2 N^2 }{64 \beta (D-1)^2} ~z^{\frac{2}{D}} ~Q^2\\
& - 48\beta k \frac{(D-1)(D-6)}{D^2 N^2} ~z^{-\frac{6}{D}} ~\dot z^2 - 36\beta k^2 z^{\frac{2(D-5)}{D}} = 0.\end{split}\ee
This is the $N$ variation equation, which does not contain second derivative term and hence is a constraint of the system under consideration. It can be easily verified that this is the Hamiltonian of the system in disguise, which reads
\be\label{Hc5iso1} \begin{split} H_c & = N \Bigg[- {1 \over 8\pi G}\left\{2{(D-1)(D-2) \over D^2N^2} z^{-{2 \over D}}\dot z^2 + 3k z^{2(D-3) \over D} - \Lambda z^{2(D-1) \over D}\right\} - \frac{\dot Q \dot z}{N} - \frac{\dot N}{N^2} Q \dot z\\
& + \frac{D^2 N^2 }{64 \beta (D-1)^2} ~z^{\frac{2}{D}} ~Q^2 - 48\beta k \frac{(D-1)(D-6)}{D^2 N^2} ~z^{-\frac{6}{D}} ~\dot z^2 - 36\beta k^2 z^{\frac{2(D-5)}{D}}\Bigg]\end{split}\ee
and is constrained to vanish. Now, using the following expression, found from the definition of momenta (\ref{pz5})
\be p_Q p_z = {(D-1)(D-2) \over 2\pi G D^2N} z^{-{2 \over D}}\dot z^2 + \dot z\dot Q + \frac{\dot N}{N}\dot z Q + 96\beta k \frac{(D-1)(D-6)}{D^2 N} ~z^{-\frac{6}{D}} ~\dot z^2,\ee
the Hamiltonian constraint equation in terms of the phase space variables is obtained as,
\be\label{Hc5iso2}\begin{split} H_c & = {1 \over 8\pi G}\left\{2{(D-1)(D-2) \over D^2N} z^{-{2 \over D}}{p_Q}^2 - 3k N z^{2(D-3) \over D} + \Lambda N z^{2(D-1) \over D}\right\} - p_Q p_z + \frac{D^2 N^3 }{64 \beta (D-1)^2} ~z^{\frac{2}{D}} ~Q^2\\
& + 48\beta k \frac{(D-1)(D-6)}{D^2 N} ~z^{-\frac{6}{D}} ~\dot z^2 - 36\beta k^2 N z^{\frac{2(D-5)}{D}} = 0 .\end{split}\ee
Finally, to express $H_c = N{\mathcal H}$, it is required to translate $H_c$ in terms of the basic variable instead of the auxiliary one. For this purpose, we choose
\be x = \frac{\dot z}{N}\ee
and therefore replace $Q$ and $p_Q$ by,
\be\label{rel} Q = \frac{\partial A}{\partial \ddot z} = \frac{\partial A}{\partial \dot x}\frac{d\dot x}{d\ddot{z}} = \frac{p_x}{N}\;\;\text{and}\;\;p_{Q} = -\dot z = -N x .\ee
In the process, one can express the Hamiltonian constraint equation (\ref{Hc5iso2}) in the following form,
\be\label{Hc5iso3}\begin{split} H_c &= N\Bigg[x p_z + {1 \over 8\pi G}\left\{2{(D-1)(D-2) \over D^2} z^{-{2 \over D}}x^2 - 3k z^{2(D-3) \over D} + \Lambda z^{2(D-1) \over D}\right\}+ \frac{D^2 }{64 \beta (D-1)^2} ~z^{\frac{2}{D}} ~{p_x}^2\\
& + 48\beta k \frac{(D-1)(D-6)}{D^2 } ~z^{-\frac{6}{D}} ~x^2 - 36\beta k^2 z^{\frac{2(D-5)}{D}}\Bigg] = N{\mathcal{H}} = 0.\end{split}\ee
It is now straight forward to express the action (\ref{A4b}) in the following canonical form
\be A = \int\left(\dot z p_z + \dot x p_x - N\mathcal{H}\right)dt~ d^{D-1} x.\ee
Corresponding quantum version is,
\be\label{qeq}\begin{split}  i\hbar z^{-\frac{2}{D}}\frac{\partial \Psi}{\partial z} &= -\frac{\hbar^2 D^2}{64 \beta x (D-1)^2}\left(\frac{\partial^2}{\partial x^2} + \frac{n}{x}\frac{\partial}{\partial x}\right)\Psi + \Bigg[{1 \over 8\pi G}\Bigg\{2{(D-1)(D-2) \over D^2} z^{-{4 \over D}}x - \frac{3k}{x} z^{2(D-4) \over D}\\
& + \frac{\Lambda}{x} z^{2(D-2) \over D}\Bigg\} + 48\beta k \frac{(D-1)(D-6)}{D^2 } ~x ~ z^{-\frac{8}{D}} - \frac{36\beta k^2}{x} ~z^{\frac{2(D-6)}{D}}\Bigg]\Psi ,\end{split}\ee
where $n$ is the factor ordering index. Again under a further change of variable, the above equation takes the look of the Schr\"odinger equation, namely
\be\label{qef}\begin{split}  i\hbar\frac{\partial \Psi}{\partial \alpha} & = -\frac{\hbar^2 D^3}{64 \beta x (D-1)^2 (D+2)}\left(\frac{\partial^2}{\partial x^2} + \frac{n}{x}\frac{\partial}{\partial x}\right)\Psi + {1 \over 8\pi G}\Bigg\{2x{(D-1)(D-2) \over D(D+2)}\alpha^{-\frac{4}{D+2}} - \frac{3k D}{x(D+2)} \alpha^{2(D-4) \over (D+2)}\\
& + \frac{\Lambda D}{x(D+2)} \alpha^{2(D-2) \over (D+2)}\Bigg\}\Psi + 12\beta k \left\{4 \frac{(D-1)(D-6)}{D (D+2) } ~x ~\alpha^{-\frac{8}{D+2}} - \frac{3 k D}{x (D+2)} ~\alpha^{\frac{2(D-6)}{D+2}}\right\}\Psi = \hat H_e\Psi,\end{split}\ee
where, $\alpha = z^{\frac{(D+2)}{D}} = a^{\frac{(D+2)}{2}}$ plays the role of internal time parameter. Note
that the effective Hamiltonian
\be \hat H_e(x, \alpha) = -\frac{\hbar^2 D^3}{64 \beta x (D-1)^2 (D+2)}\left(\frac{\partial^2}{\partial x^2} + \frac{n}{x}\frac{\partial}{\partial x}\right) + V_e(x, \alpha),\ee
is hermitian, where the effective potential $V_e$, given by,
\be\label{Ve}\begin{split} V_e(x, \alpha) &= {1 \over 8\pi G}\Bigg\{2{(D-1)(D-2) \over D(D+2)} ~x ~\alpha^{-\frac{4}{D+2}} - \frac{3k D}{x(D+2)} \alpha^{2(D-4) \over (D+2)} + \frac{\Lambda D}{x(D+2)} \alpha^{2(D-2) \over (D+2)}\Bigg\}\\
& + 12\beta k \left\{4 \frac{(D-1)(D-6)}{D (D+2) } ~x ~\alpha^{-\frac{8}{D+2}} - \frac{3 k D}{x (D+2)} ~\alpha^{\frac{2(D-6)}{D+2}}\right\},\end{split}\ee
is a function of both the so-called time variables $\alpha$ and $x$. The hermiticity of the effective Hamiltonian allows one to write the continuity equation for a particular choice of operator ordering index $n = -1$, as,
\be \frac{\partial\rho}{\partial z} + \nabla . {\bf{J}} = 0, \ee
where, $ \rho = \Psi^*\Psi ~~ \text{and} ~~  {\bf J} = ({\bf J}_x, 0, 0) $ are the probability density and the current density respectively, with,
\be {\bf J}_x = \frac{i \hbar D^3}{64 \beta x (D-1)^2 (D+2)}(\Psi\Psi^*_{~,x} - \Psi^*\Psi_{,x}). \ee
In the process, factor ordering index has also been fixed as $n = -1$ from physical argument. Further for $k=0$ extremization of the effective potential $V_e$ w.r.t $x$ yields a solution,
\be a = a_0 \exp {\sqrt{\frac{2 \Lambda}{(D-1)(D-2)}}~t}. \ee
Exponential solution for the extremum of the potential depicts that inflation is the generic feature of curvature squared action.
\subsection{ Classical and semiclassical solution (under WKB approximation)}
\subsubsection{Classical solution}
Under the standard gauge choice $N = 1$, any form of the Hamiltonian constraint equations (\ref{Hc5iso1}), (\ref{Hc5iso2}) or (\ref{Hc5iso3}) can be expressed in terms of the scale factor as
\be\label{clfld5}\begin{split} \frac{(D-1)(D-2)\dot a^2}{6 a^2} + \frac{k}{a^2} &= \frac{\Lambda}{3} - \frac{8\pi G \beta}{3} \Bigg[8 (D-1)^2 ~\frac{\dot a\stackrel{...}{a}}{a^2} - 4 (D-1)^2 ~\frac{\ddot a^2}{a^2} + 8 (D-1)^2 (D-3) ~\frac{\dot a^2\ddot a}{a^3}\\
& + (D-1)^2 (D-2)(D-10)~\frac{\dot a^4}{a^4} + 12k (D-1)(D-6)\frac{\dot a^2}{a^4} + 36\frac{k^2}{a^4}\Bigg].\end{split}\ee
The above field equation (\ref{clfld5}) is satisfied only for ($D=4$) by the following set of solutions:
\be \label{clsol1} a = \mathrm H^{-1} \cosh{( {\mathrm H} t)}, ~~ k=+1,\ee
\be \label{clsol} a = a_0\exp{( {\mathrm H} t)}, ~~~~~~ k=0,\ee
\be \label{clsol3} a = \mathrm H^{-1} \sinh{( {\mathrm H} t)}, ~~ k=-1,\ee
where, ${\mathrm H}$ is a constant. However, solution (\ref{clsol}) is a general solution of field equation (\ref{clfld5}) in arbitrary dimension provided
\be (D-3) + 16\pi G \beta H^2 (D-4)(D-1)^2 = \frac{2}{D} \left(\frac{\Lambda}{H^2} - 1\right).\ee
Therefore semiclassical approximation in $D$ dimension is possible only around the classical solution corresponding to $k = 0$. It is interesting to note that extremum of the effective potential (35) match the general solution (\ref{clsol}) for $k=0$. In fact since $\delta R|_{\partial {\mathcal V}} = 0$, only de-Sitter is allowed for higher order theory of gravity.
\subsubsection{Semiclassical solution under WKB approximation}
Instead of considering the time-dependent Schr¨odinger equation (\ref{qef}), let us, for the sake of simplicity, take up the time-independent equation (\ref{qeq}) for presenting a semiclassical solution in the standard WKB method, expressing it as
\be\label{semi} -\frac{\hbar^2 D^2 z^{\frac{2}{D}}}{64 \beta (D-1)^2}\left(\frac{\partial^2}{\partial x^2} + \frac{n}{x}\frac{\partial}{\partial x}\right)\Psi  - i\hbar x\frac{\partial \Psi}{\partial z} + V\psi = 0,\ee
where
\be V = {1 \over 8\pi G}\left\{2{(D-1)(D-2) \over D^2} z^{-{2 \over D}}x^2 - 3k z^{2(D-3) \over D} + \Lambda z^{2(D-1) \over D}\right\} + 48\beta k \frac{(D-1)(D-6)}{D^2 } ~x^2 ~z^{-\frac{6}{D}} - 36\beta k^2 ~z^{\frac{2(D-5)}{D}}.\ee
The above equation may be treated as time independent Schr{\"o}dinger equation with two variables $x$ and $z$ and therefore, as usual, let us sought the solution of equation (\ref{semi}) as,
\be\label{psisemi} \psi = \psi_0e^{\frac{i}{\hbar}S(x,z)}\ee

\noindent
and expand $S$ in power series of $\hbar$ as,
\be\label{S0} S = S_0(x,z) + \hbar S_1(x,z) + \hbar^2S_2(x,z) + .... \ .\ee

\noindent
Now inserting the expressions (\ref{psisemi}) and (\ref{S0}) in equation (\ref{semi}) and equating the coefficients of different powers of $\hbar$ to zero, one obtains the following set of equations (upto second order)
\begin{eqnarray}
 \label{hbar0} \frac{D^2 z^{\frac{2}{D}}}{64 \beta (D-1)^2}S_{0,x}^2 + x S_{0,z} + V(x,z)& = & 0, \\
 \label{hbar1}-\frac{ D^2 z^{\frac{2}{D}}}{64 \beta (D-1)^2}\left[ i S_{0,xx} - 2S_{0,x}S_{1,x} + \frac{i}{x}n S_{0,x} \right] + x S_{1,z} & = & 0,\\
 -\frac{ D^2 z^{\frac{2}{D}}}{64 \beta (D-1)^2}\left[ i S_{1,xx} - S_{1,x}^2 - 2S_{0,x}S_{2,x} + \frac{i}{x}n S_{1,x} \right] + x S_{2,z} & = & 0,
\end{eqnarray}
\noindent
which are to be solved successively to find $S_0(x,z),\; S_1(x,z)$ and $S_2(x,z)$ and so on. Now identifying $S_{0,x}$ as $p_x$ and $S_{0,z}$ as $p_z$, one can recover the classical Hamiltonian constraint equation ${\mathcal{H}} = 0$, given in equation (\ref{Hc5iso3}) from equation (\ref{hbar0}). Thus $S_{0}(x, z)$ can now be expressed as,
\be\label{S} S_0 = \int p_x dx + \int p_z dz, \ee
apart from a constant of integration which may be absorbed in $\psi_0$. The integrals in the above expression can be evaluated using the classical solution for $k = 0$ presented in equation (\ref{clsol}). This may be accomplished in view of the definition of $p_z$ given in (\ref{pz5}) along with the relation $p_x = Q$, where the expression for $Q$ is given in (\ref{Q4}) and remembering the relation (\ref{rel}), viz., $x = \dot z$ in the gauge $N = 1$. Since the probability interpretation holds only for $n = -1$, so for the semiclassical approximation we consider $k = 0$ and $n = -1$. Using the solution (\ref{clsol}), $x (= \dot z)$, $p_x$ and $p_z$ can be expressed in terms of $z$ as,
\be\label{comb} x = \frac{D}{2}{\mathrm H} z,\;\;p_x = 8\beta (D-1)^2 {\mathrm H}^2 z^{\frac{D-2}{D}} \ \ \text{and} \ \ p_z = -\frac{(D-1)(D-2) {\mathrm H}}{4\pi G D} z^{\frac{D-2}{D}} - 4\beta(D-1)^2(D-2) {\mathrm H}^3 z^{\frac{D-2}{D}}\ee
and hence the integrals in (\ref{S}) are evaluated as,
\be \int p_x dx = 2\beta (D-1) D^2{\mathrm H}^3 z^{\frac{2(D-1)}{D}} \ee
\be \int p_z dz = -\frac{(D-2){\mathrm H}}{8\pi G}z^{\frac{2(D-1)}{D}} - 2\beta D (D-1)(D-2) {\mathrm H}^3 z^{\frac{2(D-1)}{D}},\ee

\noindent
Explicit form of $S_0$ is therefore given by,
\be S_0 = -\frac{(D-2){\mathrm H}}{8\pi G}z^{\frac{2(D-1)}{D}} + 4\beta D (D-1) {\mathrm H}^3 z^{\frac{2(D-1)}{D}}. \ee
Hence, at this end the wave function reads
\be \psi = \psi_0 e^{\frac{i}{\hbar}\left[-\frac{(D-2){\mathrm H}}{8\pi G} + 4\beta D (D-1) {\mathrm H}^3 \right]z^{\frac{2(D-1)}{D}}}.\ee

\subsubsection{ First order approximation}
In the first order approximation, we take up equation (\ref{hbar1}) and express it as,
\be \label{1o}i S_{0,xx} - 2S_{0,x}S_{1,x} - \frac{i}{x}S_{0,x} - 16\beta (D-1)^2 {\mathrm H}^2 z^{\frac{D-2}{D}} S_{1,x} = 0, \ee
\noindent
using the relation, $S_{1,z}=S_{1,x}\frac{dx}{dz}$.
Equation (\ref{1o}) may further be rearranged as,
\be\label{S1} i\frac{p_{x,x}}{p_x} - i\frac{1}{x} = 2S_{1,x} + 16\beta (D-1)^2 {\mathrm H}^2 z^{\frac{D-2}{D}} \frac{S_{1,x}}{p_x} = 4S_{1,x}, \ee
\noindent
which under integration yields the following explicit form of $S_1$, viz.,
\be S_1 = \ln{\left( \frac{p_x}{x} \right)^{\frac{i}{4}}} + f_1(z).\ee
Again rewriting expression (\ref{S1}) in terms of $z$ using (\ref{comb}), one obtains
\be S_{1,z} = 0,\ee
and so $S_1$ turns out to be a function of $x$ only.
Hence, finally we obtain,
\be S_1 =  \ln{\left( \frac{p_x}{x} \right)^{\frac{i}{4}}},\ee
\noindent
and the wave function at this end (i.e., up to first order approximation) reads,
\be \psi = \psi_0 \left( \frac{x}{p_x} \right)^{\frac{1}{4}}\exp^{\frac{i}{\hbar}S_0} = \psi_0 \left\{\frac{D z^{\frac{2}{D}}}{16\beta (D-1)^2 {\mathrm H}}\right\}^{\frac{1}{4}} e^{\frac{i}{\hbar}\left[-\frac{(D-2){\mathrm H}}{8\pi G} + 4\beta D (D-1) {\mathrm H}^3 \right]z^{\frac{2(D-1)}{D}}}.\ee
The wave function has oscillatory behaviour as well and corresponds to the one obtained by Sanyal and his collaborators \cite{Sanyal3}, in $D = 4$ dimensions.

\section{Concluding remarks}
Expressing the action in terms of the three metric $h_{ij}$ and removing a part of the total derivative term prior to the introduction of auxiliary variable lead to canonical formulation of scalar curvature squared gravity in such a way that the action may be expressed in the ADM form, for Robertson-Walker minisuperspace metric in $4$ dimension. The quantum counterpart looks like Schr\"odinger equation, where an internal variable acts as the time parameter. Extremum of the effective potential gives inflationary solution. Further, continuity equation picks up a particular operator ordering index $n = -1$ and the standard quantum mechanical probability interpretation follows.\\

Such wonderful features of $R^2$ gravity led us to proceed for canonical formulation of action (\ref{A4}) in arbitrary higher dimension, since almost all effective higher order theories carry $R^2$ term in the action. In higher dimension, we observe that the action should be expressed in terms of a variable $z = a^{D\over 2} = {h_{ij}}^{D\over 4}$, instead of $h_{ij}$ to account for the boundary terms which appear under the standard metric variation technique. The same scheme \cite{Sanyal3} has been followed, and all the nice features are observed in higher dimension as well. Of particular importance is that the operator ordering index has been fixed ($n = -1$) from the physical argument that continuity equation should be valid. Nevertheless the field equations satisfy Starobinsky's solutions \cite{Staro} for $k = \pm 1$ only in $D = 4$ dimensions, while the solution (43) for $k = 0$ is valid in general for arbitrary dimension $D\ge 4$. Standard WKB approximation has therefore been performed for classical solution around $k = 0$, $n = -1$ which shows oscillatory behaviour of the wavefunction, as usual.\\

\end{document}